\documentclass[fleqn,twoside]{article}
\usepackage{espcrc2,amssymb}

\usepackage{graphicx,epsfig,color,amsfonts}
\usepackage[figuresright]{rotating}

\newcommand{\unit}[1]{\,\rm#1}

\title{Compact Frontend--Electronics and Bidirectional 3.3 Gbps Optical Datalink for Fast Proportional Chamber Readout}

\author{S. L\"uders\address[IPP]{Institute for Particle Physics, ETH Zurich, 5232 Villigen PSI, Switzerland},
R. Baldinger\addressmark[IPP],
D. Baumeister\address[ASIC]{ASIC--Laboratory, Kirchhoff--Institute for Physics, 69120 Heidelberg, Germany},
K. B\"osiger\address[UNIZH]{Physics Institute, University of Zurich, 8057 Zurich, Switzerland},
R. Eichler\addressmark[IPP]\thanks{Corresponding author. Phone: +41~1~633~2018, Fax: +41~1~633~1233, e--mail: eichler@particle.phys.ethz.ch},
M. Feuerstack--Raible\addressmark[ASIC],
C. Grab\addressmark[IPP],
S. L\"ochner\addressmark[ASIC],
B. Meier\addressmark[IPP],
P. Robmann\addressmark[UNIZH],
B.A. Schmid\addressmark[UNIZH],
U. Stange\addressmark[ASIC],
S. Steiner\addressmark[UNIZH],
U. Straumann\addressmark[UNIZH],
S. Streuli\addressmark[IPP],
K. Szeker\addressmark[IPP]\thanks{Present address: Physics Institute, University of Zurich, 8057 Zurich, Switzerland}
P. Tru\"ol\addressmark[UNIZH]
}
       
\begin{document}

\begin{abstract}
The 9600 channels of the multi--wire proportional chamber of the H1 experiment at HERA have to be read out within $96\unit{ns}$ and made available to the trigger system. The tight spatial conditions at the rear end flange require a compact bidirectional readout electronics with minimal power consumption and dead material.\\
A solution using 40 identical optical link modules, each transferring the trigger information with a physical rate of $4\times 832\unit{Mbps}$ via optical fibers, has been developed and commisioned. The analog pulses from the chamber can be monitored and the synchronization to the global HERA clock signal is ensured.
\vspace{1pc}
\end{abstract}

\maketitle
\setcounter{footnote}{0}

\section{Introduction}
An optical link and frontend electronics has been developed to read out all 9600 channels of the H1 experiment's central inner multi--wire proportional chamber ({\it CIP\/}) within the time between two bunch crossings, i.e. $96\unit{ns}$. The application requires a bidirectional multi--purpose link: the digitized chamber information has to be provided to the trigger system $40\unit{m}$ away, selected analog pulses should be accessible for monitoring purposes and the whole frontend electronics must be synchronized to the global HERA clock signal. Furthermore, the optical link and readout electronics must fit in the available space of a $130\unit{mm}$ long open cylinder with inner and outer radii of $152\unit{mm}$ and $198\unit{mm}$. No commercial solution for optical links fulfill these requirements in one compact unit.

The custom--made solution is composed of forty identical {\it optical link modules}, where a 64--fold multiplexing reduces the number of data lines. Each module performs an optical transmission with a physical rate of $3.3\unit{Gbps}$. Precisely aligned VCSEL and PIN diode arrays allow for bidirectionality. The bending of optical fibers by $90^\circ$ within $2\unit{mm}$ minimizes the overall height of the design. 

A short overview of the CIP upgrade is given in Section \ref{sec.layout} and the general layout is discussed: Each of the optical link modules consists of an on--detector unit, two optical hybrids with optical cables and a receiver unit. Their functional designs are presented in Sections \ref{sec.ondetector}, \ref{sec.hybrid} and \ref{sec.receiver}, respectively. The performance of the optical link and frontend electronics is presented in Section~\ref{sec.performance}.

\section{CIP Upgrade and General Layout}
\label{sec.layout}
With the year 2000 upgrade of the HERA electron--proton collider at DESY, an increase in luminosity by a factor of five is anticipated. The expected higher background rate, predominantly beam--wall and beam--gas reactions, necessitates an improvement of the CIP to provide high background rejection efficiency of the $z$--vertex trigger \cite{det.h1,det.cip}.

The redesigned CIP \cite{prc.cip2000} is built of five concentrical cylinders ({\it layers\/}) with radii from $152\unit{mm}$ to $198\unit{mm}$. In the azimuthal angle, each layer is equally subdivided into 16 {\it segments}, each consisting of 120 {\it pads}\footnote{In fact, the CIP uses a projective geometry requiring 119 pads on the innermost layer, and 112, 106, 99 and 93 pads on the following layers, respectively. But for symmetry reasons, each optical link module will be capable of handling 120 pads.} along the symmetry axis ({\it $z$--axis\/}). Charged particles traversing the chamber are detected by the 9600 pads, which provide space points and timing information. The direction of tracks can be inferred from the pattern of hit pads. For electron--proton collisions intersections of tracks with the $z$--axis come mostly from the interaction region, while the dominating background originates from proton beam losses upstream of the experiment. Experience has shown, that such background tracks seen by the H1 experiment are typically intersecting the $z$--axis at $0.5$ to $1.5\unit{m}$ upstream of the mean $ep$ interaction position. A new trigger system based on the latest FPGA family will identify these upstream events and is presently being commissioned. According to simulations this new trigger will improve the background rejection capability of the first level trigger by an order of magnitude compare to the previous {\it $z$--vertex trigger} \cite{det.zvtxtrigger}.

The decision has to be made for {\it every} bunch crossing. Thus all 9600 pads have to be read out within the time between two bunch crossings, i.e. within $96\unit{ns}$, corresponding to the bunch crossing frequency of $10.4\unit{MHz}$ ({\it HERA clock}). From the timing information, the bunch crossing number can be deduced.

Each of the forty identical optical link modules is used to read out all $2\times 120$ pads of two adjacent segments (a {\it double--segment\/}) of a layer with a rate of $10.4\unit{MHz}$. The {\it on--detector electronics unit} amplifies and shapes the signals from the pads, discriminates and serializes them to $4\times 15\unit{bit}$ words. After a second level 16--fold multiplexing, this {\it trigger information} is transferred to the {\it receiver electronics unit} $40\unit{m}$ away, located outside the main detector. Thus the total digitized information per module sums up to a data rate\footnote{In principle, each optical link module is capable to transmit at a data rate of $4\times 1000\unit{Mbps}$ at maximum.} of $4\times 624\unit{Mbps}$. The on--detector electronics component must be synchronized to the global HERA clock signal therefore the system requires bidirectionality. The receiver electronics provides the global HERA clock and retrieves the multiplexed trigger information, which is de--serialized and distributed to the {\it trigger system}. Additionally, analog signals from each pad are transmitted and accessible for monitoring purposes.

To retain high geometrical acceptances for the new CIP and neighbouring detectors in the H1 experi\-ment, the available space for mechanical support structures and electronics is limited to a $130\unit{mm}$ long open cylinder with inner and outer radii of $152\unit{mm}$ and $198\unit{mm}$, respectively, located at the backward end flange\footnote{The term ``backward'' labels the $-z$ end of the H1 experi\-ment pointing in the direction of the electron beam.} of the CIP.
\begin{figure}[!bt]
  \centering
  \includegraphics*[width=\linewidth,bbllx=80,bblly=210,bburx=300,bbury=540,clip=true]{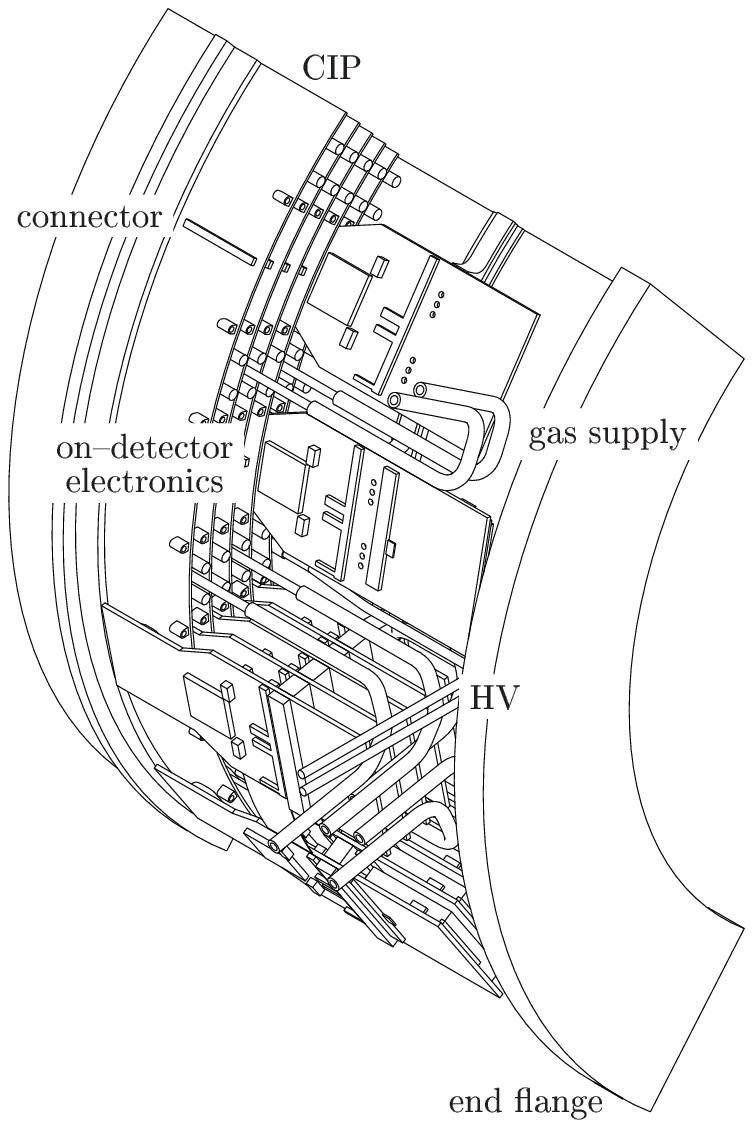}
  \caption[]{\label{det.cipendflange} Illustration of the on--detector units mounted on the CIP. Four segments of the five layer CIP with gas tubes and HV cables are shown on the left. Five on--detector units stacked on top of each other and supported by cooling blocks are plugged on the lower segments; one unit is mounted on each of the upper segments. Their top sides face to the symmetry axis.}
\end{figure}
This tight space has to be shared between on--detector electronics, their suspension and cooling, low and high voltage power cables and gas supply lines (Figure \ref{det.cipendflange}). The power consumption has to be minimized to avoid an excessive heat dissipation inside the H1 experiment due to limited cooling possibilities.

Only an optical transmission allows high serial data rates, suppresses crosstalk and decouples detector and trigger system, while reducing the number of cables and the power consumption to a minimum. A readout with copper cables as formerly done would increase that volume by a factor of ten, would require even more driving power and would produce an unwanted high contribution to the dead material. In addition, an electrical transmission at the required high rates will likely induce noise into the very sensitive liquid argon calorimeter of the H1 experiment.

The optical link between on--detector electronics and the receiver electronics is established by two {\it optical hybrids} which perform opto--electrical (re)conversion of all data lines, i.e. four digital channels for the trigger information, two analog channels and two channels for the HERA clock signal. Experience with opto--mechanics has already been collected in the ETH Zurich group. An analog optical readout for the H1 experiment has been successfully operated since 1995 \cite{det.cst1,det.cst2}.

\section{On--Detector Electronics Unit}
\label{sec.ondetector}
\begin{figure}[!b]
  \centering
  \includegraphics[width=\linewidth,clip=true]{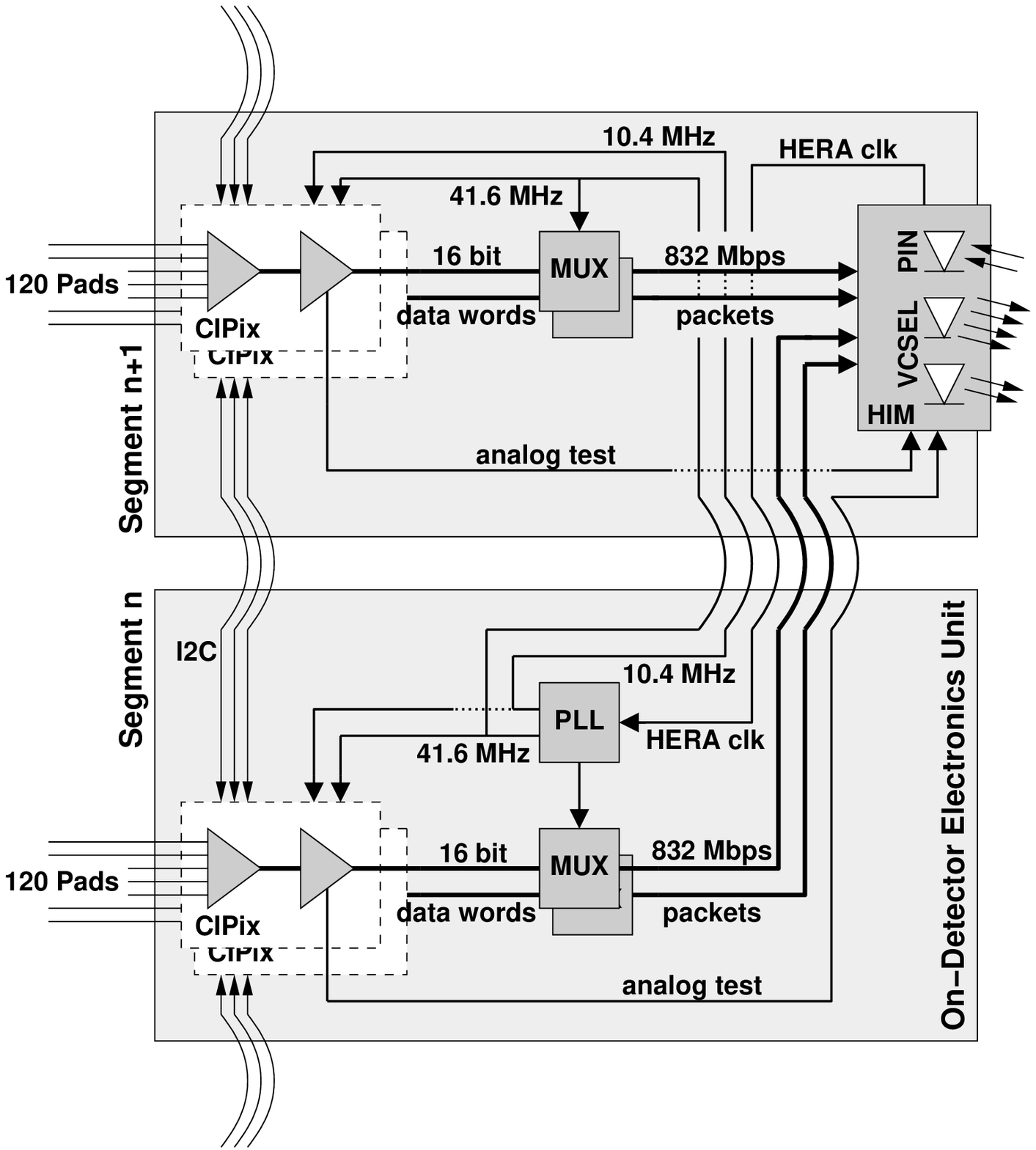}
  \caption[]{\label{fig.ondetector.flow} Signal flow of the on--detector electronics unit. A data word consists of the $15\unit{bit}$ trigger information plus the FirstWord bit. For details see text.}
\end{figure}

Each on--detector unit collects the charge from $2\times 120$ pads. Their charge is conducted via micro coax cables \cite{part.coax} to the rear end flange of the CIP. For each segment, a connector (Fujitsu FCN\,298 \cite{part.fujitsu}, 120 contacts, $500\unit{\mu m}$ pitch, $2.5\unit{mm}$ height) passes the 120 channel pad information to a pair of analog readout ({\it CIPix\/}) chips. The CIPix chip amplifies, shapes, discriminates and four--fold multiplexes the signals. Further compression is done by two 16--fold multiplexers, each driving a differential high--speed data channel at $832\unit{Mbps}$. After electro--optical conversion by the optical hybrid, the light pulses are transmitted to the receiver electronics unit. The overall synchronization is done with the global HERA clock signal received by the optical hybrid. A low jitter phase--locked--loop ({\it PLL\/}) unit \cite{part.idt} generates a $41.6\unit{MHz}$ clock signal, which is distributed to the multiplexer and --- in addition to the HERA clock signal --- to the CIPix chip. Furthermore, selected analog signals can be branched off before entering the CIPix discriminator to monitor the CIP. These {\it analog test signals} are also transmitted (Figure \ref{fig.ondetector.flow}).

\begin{figure}[!b]
  \centering
  \includegraphics*[width=\linewidth,bbllx=80,bblly=250,bburx=300,bbury=510,clip=true]{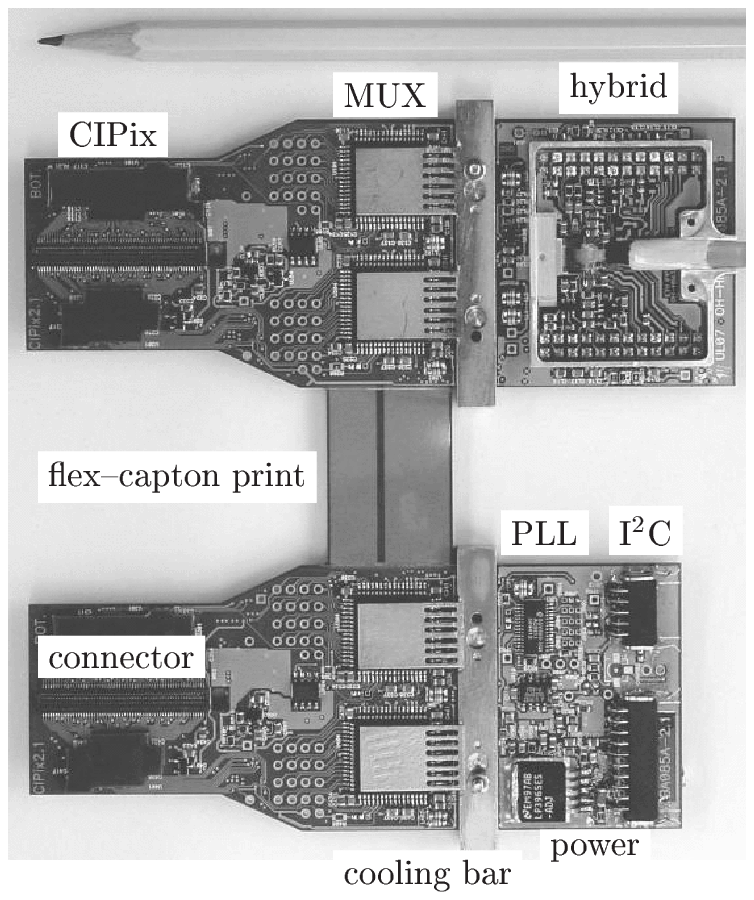}
  \caption[]{\label{pic.cipixcard} On--detector electronics rigid--flex print serving one double--segment.}
\end{figure}

For compactness, one optical hybrid serves a double--segment and is mounted on one of the two separate halves of the on--detector unit. The other half holds the PLL unit and houses the voltage regulators. Due to the curvature of the CIP, a thin four layer flex--capton print bridges all signals via striplines to their destinations: The high--speed data channels and analog test signals are transferred from the other segment to the optical hybrid and, in return, the HERA clock signal is provided to the PLL unit. The flex--capton print is sandwiched between each of the halves. This {\it rigid--flex} print is produced by Dunkel \& Sch\"urholz \cite{part.schuerholz}. It is implemented as an extremely high dense board with microstrip transmission lines and eight layers in total (Figure \ref{pic.cipixcard}).
Its outer dimensions are $130\unit{mm}$ in length, $2\times 49\unit{mm}$ in width and $9\unit{mm}$ in height. The open length of the capton print, i.e. the distance between both halves, increases proportional to the layer radii.

All on--detector units of one layer are connected via an I$^2$C bus daisy--chain \cite{part.i2c}. This allows for a steering of every CIPix chip from a terminal, e.g. all CIPix chips can be initialized layer--wise. In addition, it performs a one--wire serialized temperature measurement \cite{part.dallas}.

\subsection{CIPix IC}
\label{cipix}
The CIPix analog readout chip is custom--made \cite{asic}. It amplifies, shapes, discriminates and multiplexes the incoming signals \cite{part.cipix}.

\begin{figure*}[!hbt]
  \centering
  \includegraphics[width=\linewidth,clip=]{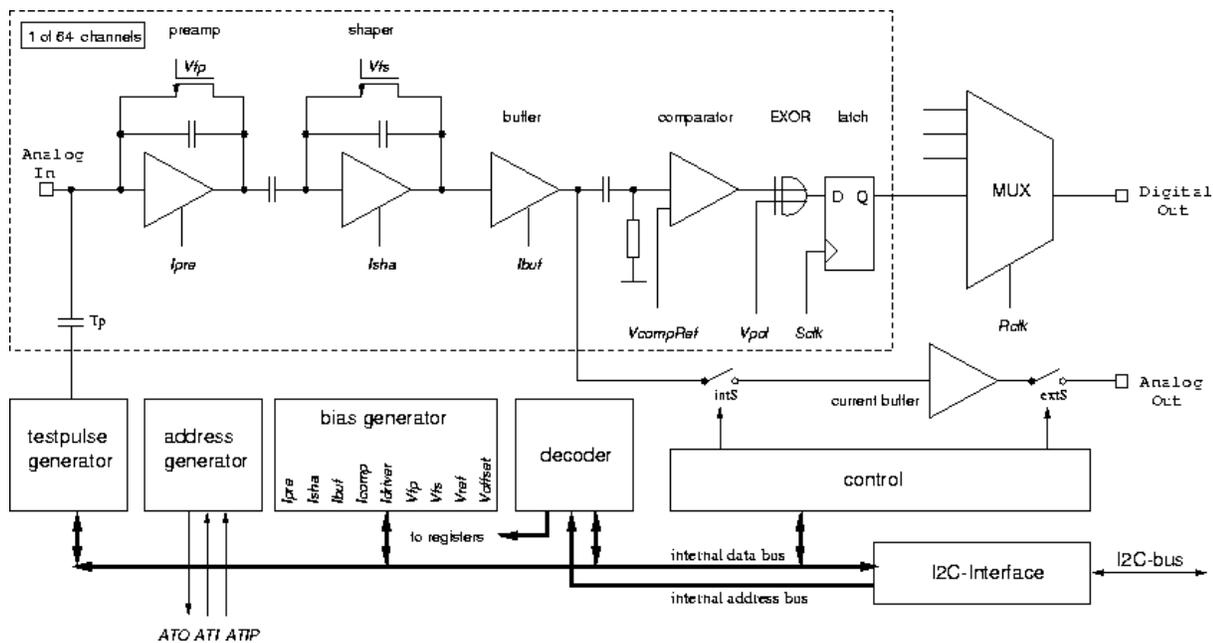}
  \caption[]{\label{fig.cipix} Block diagram of the CIPix chip.}
\end{figure*}

For each of the 64 analog input channels, it consists of a charge sensitive preamplifier with a gain of $20\unit{mV}$ per $10^5$ electrons, a CR--RC semi Gaussian shaper with a peak time of $50$ to $70\unit{ns}$ and a comparator. The discriminated signals are synchronized to the HERA clock signal and four adjacent pads are multiplexed (Figure \ref{fig.cipix}).
For monitoring and testing purposes, one of the analog signals can be selected and branched off to an analog output. Test pulses of user--defined pattern and frequency can be internally generated. Comparator thresholds, the selection of channels for analog output and test pulses are programmable via the I$^2$C bus. To protect the bond--wires and the surface, each CIPix chip is sealed (``glob--topped'') with protective glue (Epoxy Technology H70S \cite{part.epotec}).

\medskip
The analog signals from 60 pads are processed by a single CIPix chip. Synchronously to the $41.6\unit{MHz}$ clock signal, the CIPix's 15 digital output channels give four successive words with $15\unit{bits}$ each. A {\it FirstWord bit} tags the first of these words and will make it possible to maintain the synchronization to the HERA clock signal in the trigger system. Together they form the $16\unit{bit}$ {\it data word}. An EmptyDataSet signal, generated in the case of missing inputs on all 60 input pads of the CIPix chip, serves as the {\it EmptyDataSet bit}. All 160 EmptyDataSet bits provide a coarse readout and will be used for a trigger decision whether an event is compatible with cosmic ray background or not.
 
\subsection{Multiplexing\,/\,Demultiplexing}
A further reduction of channels is achieved by building a differential high--speed data link for point--to--point communication between the Hewlett--Packard HDMP\,1032 transmitter and the HDMP\,1034 receiver \cite{part.hp}.

Both bipolar chips provide the transmission of a $16\unit{bit}$ TTL data word plus one flag bit with a serial data rate of $221$ to $1190\unit{Mbps}$. The rate is chosen by an external reference clock signal synchronous with the incoming word. The HDMP\,1032 PLL\,/\,clock generator locks onto the reference clock signal and multiplies it up to the high--speed serial clock signal. From the data word a special four bit encoding information is generated on the fly, proceeding the data word. Together they give a $20\unit{bit}$ {\it packet}. On the one hand, the encoding ensures the DC balance of the serial line. The disparity\footnote{The disparity is defined as the total number of ``high'' bits minus the total number of ``low'' bits.} of each data word is determined. Depending on the disparity of the previous data word, an inversion of the actual word is done to keep a $50\unit{\%}$ duty cycle. Additionally, the encoding bits provide an error detection, tagging wrongly transmitted words, and include an user--controlled {\it flag bit}. On the other hand, the unique bit pattern of the encoding scheme incorporates the high--speed clock signal and thus saves an additional clock signal line between transmitter and receiver. The packet is serialized and leaves the HDMP\,1032 as a differential $100\unit{\Omega}$ terminated ECL compatible high--speed signal.

The HDMP\,1034 receiver's Clock Data Recovery unit separates encoding bits and data word from the $20\unit{bit}$ packet, extracts the high--speed serial clock signal and locks to its phase. The data words are (eventually) inverted and then demultiplexed. The Parallel Automatic Synchronization System synchronizes these words to an external reference clock signal. A master--slave mode allows for a synchronization of several HDMP\,1034 chips: A deviation of the relative phase of the data word and the reference clock signal generates a shift request, passed to the master. The master controls the delay of all outputs of all slaves.

\medskip
Each $16\unit{bit}$ data word is serialized by a HDMP\,1032 chip. The EmptyDataSet bit is used as input for the flag bit; the $41.6\unit{MHz}$ clock signal provides the reference clock signal for both HDMP chips. Therefore the digitized trigger information from 60 pads is transmitted every $96\unit{ns}$ with a data rate of $624\unit{Mbps}$. The overall physical rate includes in addition per $96\unit{ns}$ the four FirstWord bits and $4\times 4$ bits of the encoding scheme (with the EmptyDataSet bit) and amounts to $832\unit{Mbps}$. The receiver chip re--parallelizes the trigger information and extracts the FirstWord bit and EmptyDataSet bit. Four transmitter\,/\,receiver chip pairs --- one for each CIPix chip --- are used per module. 

\begin{figure}[!bt]
  \centering
  \includegraphics[width=\linewidth,clip=true]{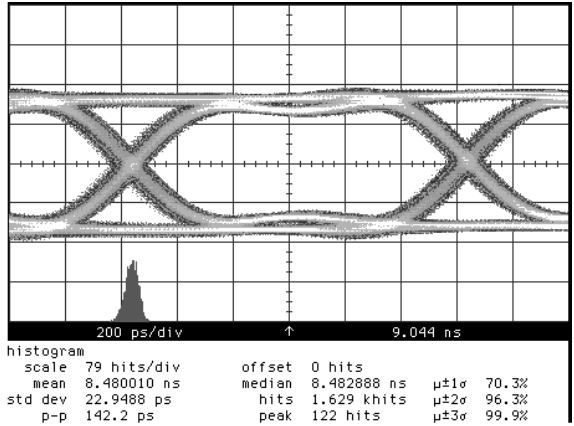}
  \caption[]{\label{pic.digieye1} Eye--diagram accumulated from one digital channel and measured directly at the differential output of the 16--fold multiplexer. Shown is the digital level (``high''\,/\,''low'') superimposed for all bits as a function of time. The small histogram gives the jitter. The signal density is given by the greyscale: the lighter the denser.}
\end{figure}

Bits of one high--speed digital channel have been superimposed for the eye--diagram (Figure \ref{pic.digieye1}).
The rising and falling edges are well separated. The measurement of the zero--crossing of the rising edge results in a jitter of the data words of $23\unit{ps}$ before entering the optical hybrid.

\section{Optical Hybrid}
\label{sec.hybrid}
The optical hybrids constitute the interface between the electrical and the optical r\'egime.

Following the signal flow from the CIP to the trigger system, the optical hybrid on the on--detector unit ({\it HIM}, High--Speed Interconnection Module) acts as a driver for the outgoing $20\unit{bit}$ packets and analog test signals and as a receiver for the incoming HERA clock signal. The optical hybrid on the receiver unit side ({\it DeHIM\/}) acts vice versa. The interfacing pins require or deliver differential CMOS logic signals for each data channel, respectively.

Each HIM\,/\,DeHIM pair serves one double--seg\-ment, i.e. transmits four multiplexed data channels with a physical rate of $4\times 832\unit{Mbps}$, two analog test signal channels and two HERA clock signal channels via an optical fiber array with eight fibers.

The four data channels are driven by the Helix HXT\,2000 \cite{part.helix} chip optimized for vertical cavity surface emitting laser ({\it VCSEL\/}) diodes. The VCSEL diodes convert the electrical signal to light pulses. After $40\unit{m}$ of optical fibers, conventional PIN diodes reconvert the optical to electrical signals. These are amplified by the Helix HXR\,2004 receiver chip and produce four differential data signals. Because of the high data rates, the hybrid boards are impedance controlled and realized in four--layers with layer--to--layer blind--via connections \cite{part.schuerholz}.

For redundancy two HERA clock signals are transmitted in parallel. Standard SZ\,125 drivers match their signal levels with respect to the VCSEL diode specifications. Conexant (formerly Microcosm) MC\,2007 receivers \cite{part.conexant} convert the PIN diode responses back to voltage--modulated signals. Its active gain control ({\it AGC\/}) ensures a stable output signal above the sensitivity limit at about $-20\unit{dBm}$. The two analog test signals are driven by Maxim MAX\,4212 operation amplifiers \cite{part.maxim} and received by Conexant MC\,2011 (without AGC) chips. These differential analog signals and the HERA clock signal are finally amplified by Maxim MAX\,4212 chips. 

\subsection{Driver and Receiver Circuits}
The Helix HXT\,2000 driver and the HXR\,2004 receiver chips are designed for high--speed optical transmission up to $1.25\unit{Gbps}$ per channel; the HXT\,2000 is optimized for VCSEL diodes at $800$ to $1500\unit{nm}$ wavelength. The differential inputs to the HXT\,2000 --- four of them enabled --- are amplified and current--modulated. External resistors allow to control the average and modulation current collectively for a VCSEL diode array. Thus the working range, i.e. the laser current, of an array of four VCSEL diodes can be optimized for maximum optical output.

The four channel HXR\,2004, compatible to $0.6\unit{pF}$ photodiode arrays, converts the photocurrent from the PIN diodes to differential output voltages. In addition, the average photocurrents can be monitored.

\subsection{Vertical Cavity Surface Emitting Laser Diodes}
Laser diodes convert current--modulated signals into power--modulated light signals.

\begin{figure}[!b]
  \centering
  \includegraphics[width=\linewidth,clip=]{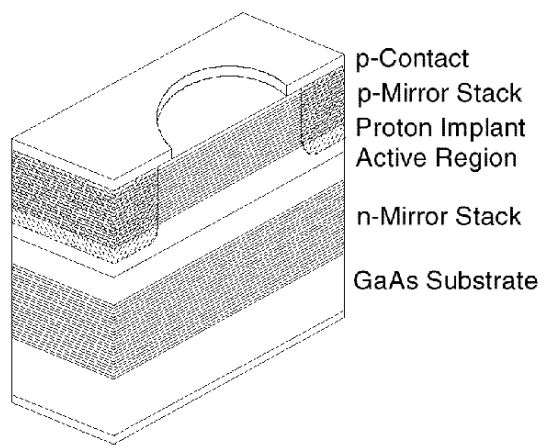}
  \caption[]{\label{fig.vcsel} Sketch of a VCSEL diode.}
\end{figure}

In a VCSEL diode the light propagates vertically through the structure (Figure \ref{fig.vcsel}).
With this orientation the laser cavity can be grown to match the wavelength of the laser light, i.e. $850\unit{nm}$. The total spectral width of the emission is generally less than $0.5\unit{nm}$, which ensures a low coherence source. The beam divergence is typically below $12^\circ$ FWHM. Alternating layers of AlAs and Al$_{0.15}$Ga$_{0.85}$As provide the p-- and. n--mirror stack surrounding the active region, respectively. With contact to the p--mirror, the VCSEL anode is bonded to the modulating current line, while the GaAs substrate holds the cathode i.e. ground potential \cite{part.honeywell}.
Both, VCSEL and PIN diodes, are grouped in six-- and two--diodes dies, respectively, from Truelight Corporation \cite{part.truelight}. The specifications for the VCSEL and PIN diode arrays used are given in Table \ref{tab.vcsel}.
\begin{table*}[!bt]
  \caption[]{\label{tab.vcsel} Specifications of the VCSEL and PIN diode arrays.}
  \renewcommand{\arraystretch}{1.2} 
  \centering
  \begin{tabular}{lcc}
    \hline
    \hline
    \multicolumn{1}{l}{\rule[-3mm]{0mm}{8mm}{Optical specifications:}}&VCSEL&PIN\\
    \hline
    \hline
    Wavelength&850\,nm, multimode&$850\unit{nm}$\\
    Beam profile and divergence&round, $<12^\circ$&\\
    Active\,/\,Sense area diameter&$18\unit{\mu m}$&$120\unit{\mu m}$\\ 
    Slope efficiency\,/\,Response at $850\unit{nm}$&$>0.125\unit{W/A}$&$>0.6\unit{A/W}\pm 3\unit{\%}$\\
    \hline
    \hline
    \multicolumn{1}{l}{\rule[-3mm]{0mm}{8mm}{Electrical specifications:}}&VCSEL&PIN\\
    \hline
    \hline
    $U_{\rm operating}$&$1.7..2.3\unit{V}$&\\
    $U_{\rm reverse}$&$>10\unit{V}$&$>10\unit{V}$\\
    Serial impedance&typ. $30\unit{\Omega}$&\\
    $I_{\rm laser}$\,/\,$I_{\rm dark}$&$4.5\unit{mA}$&$<40\unit{nA}$\\
    $C_{\rm total}$&&$<0.9\unit{pF}$ @ $5\unit{V}$\\
    $\tau_{\rm rise\,/\,fall}$&$<250\unit{ps}$&$100\unit{ps}$\\
    Crosstalk&&$>30\unit{dB}$\\
    \hline
    \hline
    \multicolumn{1}{l}{\rule[-3mm]{0mm}{8mm}{Mechanical specifications:}}&\multicolumn{2}{c}{VCSEL\,/\,PIN}\\
    \hline
    \hline
    Operating temperature&\multicolumn{2}{c}{$<85^\circ\unit{C}$}\\
    Chip thickness&\multicolumn{2}{c}{$150\unit{\mu m}$}\\
    Pitch diode\,/\,diode&\multicolumn{2}{c}{$250\unit{\mu m$}}\\
    \hline
    \hline
  \end{tabular}
\end{table*}
The VCSEL diodes are specified as ``class IIIb laser'' in the safety standard ANSI Z136.1 \cite{ansi} and have to be treated as a potential eye hazard.

\medskip
Because of a delicate passivation, the softer ball--bond process using golden bond wires has been preferred to the wedge--wedge bonding and to the use of aluminum bond wires. Thus ultrasonic vibrations acting on the VCSEL diodes could be minimized and harm to the passivation could be prevented. Together with the required specifications of at least $1.5\unit{mW}$ at $12\unit{mA}$ laser current (i.e. to a slope efficiency of $0.125\unit{W/A}$) and an uniform power gain over the VCSEL array, the yield has been tested to be about $31\unit{\%}$.

\begin{figure}[!hbt]
  \centering
  \includegraphics[width=\linewidth,clip=true]{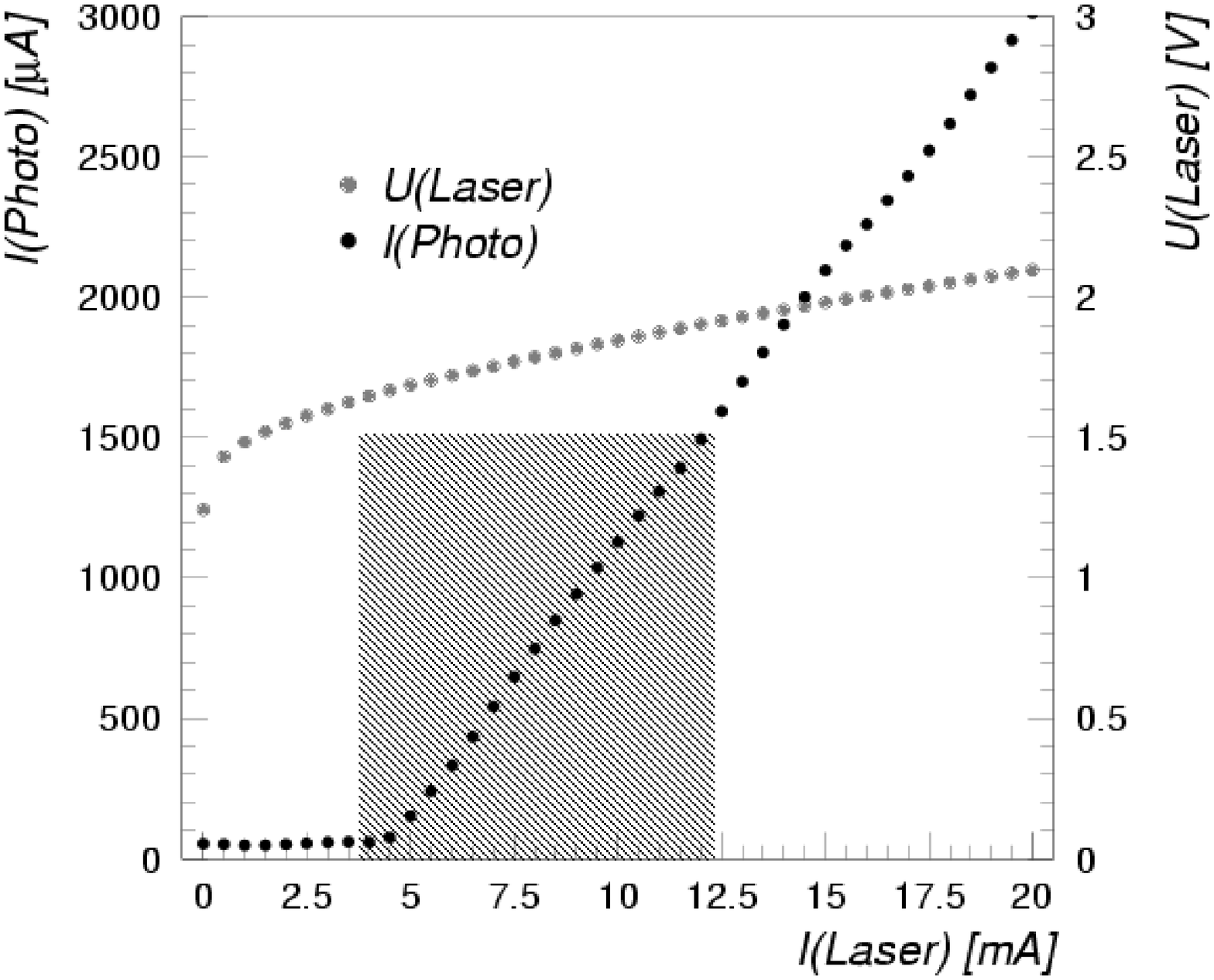}
  \caption[]{\label{fig.vcseltune} Output power ($\sim I_{\rm Photo}$) and characteristic curve ($U_{\rm Laser}$) of a VCSEL diode. The shaded area marks the working range.}
\end{figure}

The lasering of the used VCSEL diodes typically starts at a current of $4.5\unit{mA}$ (Figure \ref{fig.vcseltune}).
With appropriate settings for the average and modulation current of the HXT\,2000 ($V_{\rm avg}=1.74\pm 0.13\unit{V}$ and $V_{\rm mod}=1.38\pm 0.08\unit{V}$, respectively) the working range has been optimized. The digital signal levels vary beween $-49.1\pm 4.5\unit{dBm}$ for logical ``low'' and $-3.7\pm 0.8\unit{dBm}$ for logical ``high'', leading to a dynamic range in optical output of approximately $45\unit{dB}$. The uniformity over a VCSEL diode array is about $1.3\pm 0.7\unit{dB}$.

\subsection{\boldmath Alignment and $90^\circ$ Bending}
\begin{figure}[!b]
  \centering
  \includegraphics*[width=\linewidth,bbllx=60,bblly=270,bburx=270,bbury=480,clip=true]{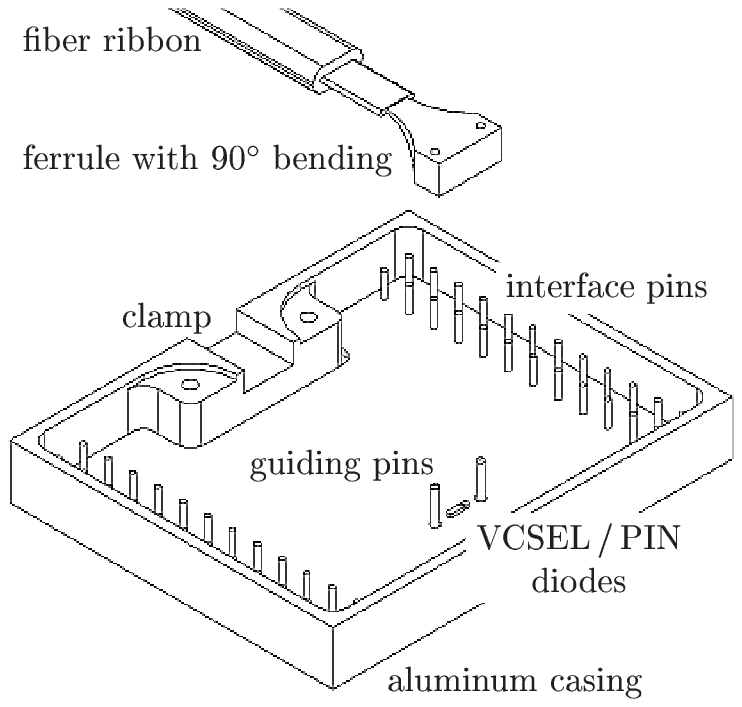}
  \caption[]{\label{fig.casing} Exploded view of the aluminum casing. Between two guiding pins for the ferrule (enclosing the $90^\circ$ fiber deflection) are the aligned sixfold VCSEL and twofold PIN diode arrays.}
\end{figure}

In case of the HIM, the sixfold VCSEL and twofold PIN diode arrays are aligned with a precision of $5\unit{\mu m}$ with respect to each other and with respect to two guiding pins \cite{part.diamond}. This provides a pitch of $250\unit{\mu m}$ in order to match the pitch of conventional fiber ribbon connectors ({\it MTP connectors} \cite{part.uscon}). The guiding pins adjust the connector to the diode arrays (Figure \ref{fig.casing}).

To obtain the desired precision, each diode array is positioned by a custom--made micro manipulator. The manipulator is mounted on an $xy$ table, which makes it possible to perform an accurate position measurement and an optical survey. At its final position, the diode array is lowered and glued onto the hybrid. An optimal mechanical and optical performance has been achieved using a two component conductive glue (Epoxy Technology H21D \cite{part.epotec}) with a resistivity of $3\cdot10^{-4}\unit{\Omega/\square}$ and a bakeout time of $2\unit{h}$. On the DeHIM side, sixfold PIN and twofold VCSEL diode dies are aligned with the same accuracy.

Since the distance between two CIP layers is less than $9\unit{mm}$, it is not feasible to mount the MTP connector above the VCSEL\,/\,PIN array. Even if the connector is reduced to its inner core, namely the {\it ferrule}, adjusting the ferrule perpendicular to the hybrid leaves no space to properly fix the connector to the optical hybrid. In addition, it complicates the installation of the fibers at the end flange. Consequently the ferrule had to be mounted parallel to the optical hybrid, i.e. parallel to the $z$--axis. Since the diodes send the light perpendicular to the die, the light needs to be redirected. Therefore the $62.5\,/\,125\unit{\mu m}$ fibers are bent within $2\unit{mm}$ of height by modifying the ferrule and by using special fibers (GGP fibers from \cite{part.3M}). The performance of the transmission line has been measured to remain stable while the attenuation at each deflection lies below $2.1\pm 0.9\unit{dB}$. On the DeHIM side, the fibers are conventionally bent within $10\unit{mm}$.

Each optical hybrid is embedded in an aluminum casing to provide robustness and handiness, to avoid electrical induction from outside and to shield the VCSEL and PIN diodes from dust. Its outer dimensions are $41\times 33\times 6\unit{mm^3}$. Clamps at the aluminum casing give a proper mechanical connection of the hybrid with the fiber tails of $600\unit{mm}$ length ($300\unit{mm}$ at the DeHIM) and prevents outside stress to affect the precise mechanical adjustment of the ferrules to the diodes.

\subsection{Optical Cables}
In guiding the optical fibers out of the main detector, several aspects have to be taken into account. Safety rules demand that all fibers and the sheathings be halogen--free and inflammable, space constraints require low bending radii of the cables, the CIP and parts of the on--detector electronics should be as easy to maintain as possible and all cables need to have a connection at the {\it cable distribution area} (CDA) at the backward end of the CIP. For these reasons, the optical transmission line is divided into four parts. Starting at the HIM, its fiber tail is plugged to a $3\unit{m}$ long fiber ribbon cable ending at the CDA. From there, a $36\unit{m}$ cable feeds the signals to the electronics cabin outside the main detector and is connected to the tail firmly attached to the DeHIM casing.

The $600\unit{mm}$ tails, including the $90^\circ$ bending part, are produced by Sch\"utten Optocommunication \cite{part.schuetten}. The long distance cables and $300\unit{mm}$ tails are made by Infineon \cite{part.infineon}; the short distance cables are from Siecor \cite{part.siecor}. The specifications of the cables are listed in Table \ref{tab.fiber}.
\begin{table}[!bt]
  \caption[]{\label{tab.fiber} Specifications of the fiber ribbons.}
  \renewcommand{\tabcolsep}{0.75pc} 
  \renewcommand{\arraystretch}{1.2} 
  \centering
  \begin{tabular}{lc}
    \hline
    \hline
    \multicolumn{2}{l}{\rule[-3mm]{0mm}{8mm}{Mechanical specifications:}}\\
    \hline
    \hline
    Fiber count\,/\,type&$12\times 62.5\,/\,125\unit{\mu m}$\\
    Flame resistance&UL-910 (Siecor)\\
                    &LSZH (Infineon)\\
    Dimensions&$<4.6\times 2.1\unit{mm^2}$\\
    Minimum bending&\\
    radius (long term)&\raisebox{1.5ex}[0pt]{$30\unit{mm}$}\\
    \hline
    \hline
    \multicolumn{2}{l}{\rule[-3mm]{0mm}{8mm}{Optical specifications:}}\\
    \hline
    \hline
    Maximum attenuation&$4.0\unit{dB/km}$\\
    Minimum bandwidth&$160\unit{MHz\times km}$\\
    Numerical aperture&$0.275\pm 0.020$\\
    \hline    
    \hline
  \end{tabular}
\end{table}
All cables are assembled with standard MTP connectors (except the ferrule end of the tails) with a typical attenuation of $0.3$ to $0.5\unit{dB}$ at each MTP--MTP connection and three connectors per link. Adapters from AMP \cite{part.amp} attach two MTP connectors to another. To obtain a predictable timing between different modules, all cables of each type are chosen to have equal length. Measurements give an average length of $3.12\pm 0.05\unit{m}$ and $36.03\pm 0.15\unit{m}$, respectively.

\section{Receiver Electronics Unit}
\label{sec.receiver}
\begin{figure}[!b]
  \centering
  \includegraphics[width=\linewidth,clip=true]{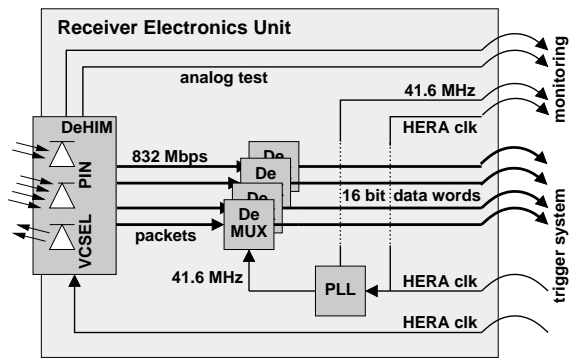}
  \caption[]{\label{fig.receiverflow} Signal flow of the receiver electronics unit.}
\end{figure}

The receiver unit provides the signals of four adjacent pads, i.e. four successive data words and the EmptyDataSet bits, to the trigger system (Figure \ref{fig.receiverflow}).

The DeHIM receives the high speed data signals and passes them to the four HDMP\,1034 demultiplexers regaining the data words. The demultiplexers are used in the master--slave daisy--chain mode to maintain synchronization of all words, at which anyone of the HDMP\,1034 can serve as master. Latches feed the four--fold multiplexed data words and the EmptyDataSet bit with a rate of $41.6\unit{MHz}$ to the backplane that connects the trigger system and the receiver electronics. The incoming global HERA clock signal is received from the backplane and directed via the DeHIM to the on--detector PLL unit. It is also passed to the receiver board's low jitter PLL unit \cite{part.idt} producing the HDMP\,1034 reference clock signal. This $41.6\unit{MHz}$ clock signal and the FirstWord bit are used in the trigger system for the synchronization of different receiver units. 

The receiver unit board is implemented as a six layer, high density and high speed board with microstrip transmission lines produced by Alwaprint \cite{part.alwaprint}. At the receiver unit's frontplate, the following signals are available for monitoring purposes: the $16\unit{bit}$ data word, the EmptyDataSet bit, the differential analog test signals, the HERA-- and the $41.6\unit{MHz}$ clock signal.

\section{Performance}
\label{sec.performance}
The optical link modules will be operated inside the H1 experiment in a $1.16\unit{T}$ magnetic field and only $15$ to $20\unit{cm}$ away from the electron beam. Thus they will be exposed to synchrotron radiation and not be accessible from outside without major effort.

Therefore, the modules needed to be tested beforehand for long--term stability, reliability and robustness. Special attention has been paid to the bit--error rate, the transmission of the analog signals and the power dissipation.

Prototypes of the modules have been successfully operated since November 1999. Neither a break--down of any of the used components nor a decrease in the power output of the VCSEL diodes have been observed.

\subsection{Link Performance}
As soon as the global HERA clock signal is applied, each link module runs autonomously. The PLL unit of the on--detector electronics locks on the HERA clock signal and distributes the HERA-- and $41.6\unit{MHz}$ clock signals to the CIPix chip and multiplexer. After $650\unit{\mu s}$ the multiplexer has been able to lock on the $41.6\unit{MHz}$ clock signal and the optical link is established. Frequency changes of the HERA clock signal within a window of $8.6$ to $11.4\unit{MHz}$ have been proven to be tolerable.

\begin{figure}[!bt]
  \centering
  \includegraphics[width=\linewidth,clip=true]{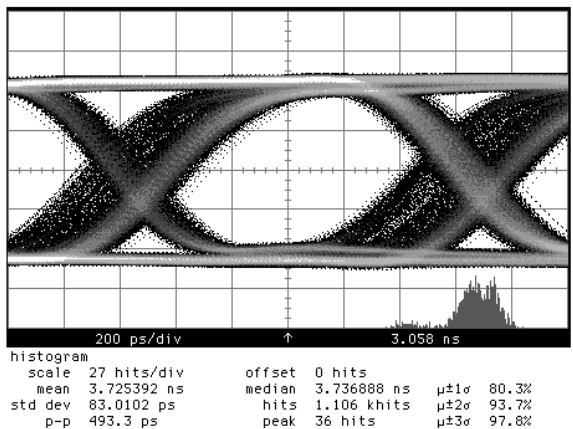}
  \caption[]{\label{pic.digieye2} Eye--diagram accumulated from one digital channel and measured directly at the demultiplexer's input. Shown is the digital level (``high''\,/\,''low'') superimposed for all bits as a function of time. The small histogram gives the jitter. The signal density is given by the greyscale: the lighter the denser.}
\end{figure}

For random data words, i.e. words with $50\unit{\%}$ duty cycle, the average light yield has been measured to be $-8.7\pm 1.8\unit{dBm}$ at the DeHIM's end of the link after $40\unit{m}$ and three connector pairs. The PIN diodes' high response of typically $63\unit{\mu A}$ drives the HXR\,2004 receiver into saturation, thus noise is suppressed. From the measurement of the bit--error--rate (see below), the lower limit has been estimated to be approximately $15\unit{\mu A}$. An eye--diagram of the digital information directly at the demultiplexer's input is shown in Figure \ref{pic.digieye2}.
The rising and falling edges are well separated. The jitter of the $20\unit{bit}$ packets is about $83\unit{ps}$. A few percent of the entries shift to earlier crossings, as can be seen in the early sideband of the jitter histogram. The crosstalk between two digital channels and the crosstalk of digital to analog channels lies below $-20\unit{dB}$. 

For quantitative tests $16\unit{bit}$ pseudo--random bit patterns were used to simulate the data words at the multiplexer's input. After transmission via the full $40\unit{m}$ link and after demultiplexing, these patterns have been compared with the original input to determine the bit--error--rate. Over a period of ten days, three errors occurred, corresponding to a bit--error--rate below $10^{-14}$. This lies far below the tolerated rate of $10^{-9}$, i.e. one error per second. All errors could be related to instabilities in the external power supply.

Problems with the synchronization between HDMP\,1032 and HDMP\,1034 have been seen if the data words imitate the bit pattern of the 16--fold multiplexer's encoding scheme for some hundred periods or if a bit next to an encoding bit is periodical in such a manner that a bit--shift results in another valid encoding bit pattern. In both cases the Clock Data Recovery unit locks on the fake bit pattern instead of the genuine encoding bits. In the operational mode of concern for the H1 experiment, this would require the same pattern of 60 pads of one segment (in a very special arrangement) repeated over many bunch--crossings. This is highly improbable and no reason for concern.

The transmission of the analog signals has been optimized to realize a one--to--one image of the chamber signals. Therefore the amplification of the CIPix chip and of the Maxim drivers in the optical hybrids have been fine--tuned. The peak--time of the analog signal is in the order of $50\unit{ns}$. From the analog signals, a delay time between CIPix chip input pads and receiver electronics unit frontend of $230\unit{ns}$ has been measured. This is dominated by the delay in the optical fibers of $200\unit{ns}$. A summary of the specifications of the optical link module is given in Table \ref{tab.ole}.
\begin{table}[!bt]
  \caption[]{\label{tab.ole}Specifications of one optical link module.}
  \renewcommand{\arraystretch}{1.2} 
  \centering
  \begin{tabular}{lr@{ @ }l}
    \hline
    \hline
    \multicolumn{3}{l}{\rule[-3mm]{0mm}{8mm}{Optical specifications:}}\\
    \hline
    \hline
    $P_{\rm out}$ (digital ``high'')&\multicolumn{2}{c}{$-3.7\pm 0.8\unit{dBm}$}\\
    $P_{\rm out}$ (digital ``low'')&\multicolumn{2}{c}{$-49.1\pm 4.5\unit{dBm}$}\\
    Dynamic range&\multicolumn{2}{c}{$45\unit{dB}$}\\
    Att. at $90^\circ$ bending&\multicolumn{2}{c}{$2.1\pm 0.9\unit{dB}$}\\
    Att. per connector&\multicolumn{2}{c}{$0.3\ldots 0.5\unit{dB}$}\\
    Crosstalk&\multicolumn{2}{c}{$<-20\unit{dB}$}\\
    $\langle P\rangle$ after $40\unit{m}$&\multicolumn{2}{c}{$-8.7\pm 1.8\unit{dBm}$}\\
    \hline
    \hline
    \multicolumn{3}{l}{\rule[-3mm]{0mm}{8mm}{Electrical specifications:}}\\
    \hline
    \hline
    Jitter HERA clock&\multicolumn{2}{c}{$43\unit{ps}$}\\
    Jitter $41.6\unit{MHz}$ clk&\multicolumn{2}{c}{$49\unit{ps}$}\\
    Jitter $20\unit{bit}$ packet&\multicolumn{2}{c}{$83\unit{ps}$}\\
                     &\multicolumn{2}{c}{(before HIM: $23\unit{ps}$)}\\
    Delay time&\multicolumn{2}{c}{$230\unit{ns}$}\\
              &\multicolumn{2}{c}{(fibers: $200\unit{ns}$)}\\
    Bit--error--rate&\multicolumn{2}{c}{$<10^{-14}$}\\
    \hline
    \multicolumn{3}{l}{\rule{0mm}{5mm}{Power Dissipation per digital channel}}\\
    &\multicolumn{2}{c}{\rule[-3mm]{0mm}{5mm}{(total 240 channels)}}\\
    \hline
    On--detector unit&$20\unit{mW}$&$+3.95\unit{V}$\\
                     &$ 4\unit{mW}$&$-3.95\unit{V}$\\
    Receiver unit    &$13\unit{mW}$&$+3.3\unit{V}$\\
    \hline
    \multicolumn{3}{l}{\rule{0mm}{5mm}{Power Dissipation per analog channel}}\\
    &\multicolumn{2}{c}{\rule[-3mm]{0mm}{5mm}{(total 2 channels)}}\\
    \hline
    On--detector unit&$110\unit{mW}$&$+4.4\unit{V}$\\
                     &$110\unit{mW}$&$-4.4\unit{V}$\\
    Receiver unit    &$135\unit{mW}$&$+5\unit{V}$\\
                     &$384\unit{mW}$&$-12\unit{V}$\\
    \hline
    \hline
  \end{tabular}
\end{table}

\subsection{Mechanical Tests}
The heat dissipation has been calculated from the measurement of the power consumption of one on--detector unit and gives $6.2\unit{W}$ per module. The total dissipation at the CIP end flange, i.e. the sum of all 40 modules, is about $250\unit{W}$, sufficiently low for a water based cooling.

The on--detector electronics has been operated in a magnetic field from $0$ to $2\unit{T}$ to simulate the impact of the H1 experiment's magnetic field. The optical output of the VCSEL diodes, the threshold of the CIPix chip, the analog pulse heights, the noise level and the total power consumption of the on--detector electronics have shown no variations within the measurable precision. The jitter of the HERA clock signal remains stable, while jitter of the $41.6\unit{MHz}$ clock signal increases from $49$ to $53\unit{ps}$ with a phase shift of $18\unit{ps}$. Thus no losses in the performance of the optical link due to the magnetic field are expected.

The irradiation at the CIP end flange and thus at the on--detector electronics is estimated to be less than $50\unit{Gy}$ per year \cite{det.cst2}. The VCSEL diodes have been irradiated to a flux of about $2\times 10^{14}\unit{{\rm neutrons}/cm^2}$, with no measurable change in either threshold or efficiency \cite{part.atlas}. After exposure to $200\unit{Gy}\pm 4\unit{\%}$  from a $^{60}$Co source, the optical fiber tails and the short distance cables have shown no change in the optical behavior. The same is expected for the $36\unit{m}$ long cables.

\subsection{Installation}
Forty optical link modules have been successfully installed at the CIP rear end flange in April 2001. All spatial requirements are met. No additional noise has been induced into the liquid argon calorimeter. All modules run autonomously with bit--error--rates well below $10^{-9}$. Due to contact problems of the chamber connectors and due to broken bonds at the CIPix chip inputs $0.8\unit{\%}$ of all channels are lost. Another $0.9\unit{\%}$ inefficiency results from the failure of two VCSEL diodes providing digital signals. For the same reason, one analog channel cannot be monitored.

\section{Summary}
A fast and compact frontend electronics has been developed to transfer $40\times 4\times 832\unit{Gbps}$ trigger information from the H1 experiment's central inner multiwire proportional chamber to the corresponding trigger system. Forty identical modules have been successfully installed at the chambers end flange and fulfill the tight spatial constraints, while the power dissipation is only about $250\unit{W}$. The optical transmission has been optimized and is functioning reliably. The bit--error--rate for each module is around $10^{-14}$.

\section*{Acknowledgments}
The optical link project has been supported by the Swiss National Science Foundation. 

\clearpage
%

%
\end{document}